\documentclass[runningheads]{llncs}
\usepackage{graphicx}
\usepackage{bbm}
\usepackage{amssymb}
\usepackage{amsmath}
\usepackage{caption}
\usepackage{subcaption}
\usepackage{multirow}
\usepackage{xurl}
\usepackage[nocompress]{cite}

\newcommand*\samethanks[1][\value{footnote}]{\footnotemark[#1]}

\begin{document}
\title{Graph-based Recommendation for Sparse and Heterogeneous User Interactions}
\titlerunning{Graph-based RS for Sparse and Heterogeneous User Interactions}
%
\author{Simone Borg Bruun\thanks{The first and second author contributed equally.}\inst{1}\orcidID{0000-0003-1619-4076} \and
Kacper Kenji Le\'{s}niak\samethanks\inst{1} 
\and
Mirko Biasini\inst{2} 
\and 
Vittorio Carmignani\inst{2} 
\and
Panagiotis Filianos\inst{2}
\and
Christina Lioma\inst{1}\orcidID{0000-0003-2600-2701}
\and
Maria Maistro\inst{1}\orcidID{0000-0002-7001-4817}}

\authorrunning{Bruun et al.}
%
\institute{Department of Computer Science, University of Copenhagen, Denmark\\
\email{\{simoneborgbruun,kkl,c.lioma,mm\}@di.ku.dk}
\and FullBrain, Copenhagen, Denmark\\
\email{\{mirko,vitto,panos\}@fullbrain.org}
}
%
\maketitle              
\begin{abstract}
Recommender system research has oftentimes focused on approaches that operate on large-scale datasets containing millions of user interactions.
However, many small businesses struggle to apply state-of-the-art models due to their very limited availability of data.
We propose a graph-based recommender model which utilizes heterogeneous interactions between users and content of different types and is able to operate well on small-scale datasets.
A genetic algorithm is used to find optimal weights that represent the strength of the relationship between users and content.
Experiments on two real-world datasets (which we make available to the research community) show promising results (up to $7\%$ improvement), in comparison with other state-of-the-art methods for low-data environments. These improvements are statistically significant and consistent across different data samples.

\end{abstract}
\section{Introduction}
\label{sec:intro}

With the advent of the internet, huge amounts of data have become available.
This allows to design and develop novel Recommender Systems (RSs) based on complex Machine Learning (ML) and Deep Learning (DL) approaches, often characterized as data-hungry approaches.
Many recent recommender models belong to this category, so a recommender dataset of size $100$K might already be considered small~\cite{Kuzelewska2020}. 
Moreover, when using such datasets, a pre-processing step is often applied to remove all users with less than a certain number of interactions, e.g., $5$, because several models are not able to learn with only few data points per user~\cite{SunEtAl2020,LudewigEtAl2021,LatifiEtAl2021,10.1145/3442381.3450011}.

In the era of big data, Small and Medium Enterprises (SMEs) struggle to find their way, given that they might not have access to such a huge amount of data.
However, SMEs are fundamental actors in the global economy, as they represent about $90\%$ of businesses and more than $50\%$ of employment worldwide~\cite{WorldBankSME}.
In these cases, RSs able to cope with low data scenarios are necessary~\cite{conf/edm/HansenHHAL17}.

In ML and DL, small data problems are notoriously hard and are usually solved with a number of well-studied techniques~\cite{SmallDataProblems}: 
(1) data augmentation, where synthetic samples are generated from the training set~\cite{LeeEtAl2009,WangQinyongEtAl2019,XiaEtAl2022};
(2) transfer learning, where models learn from a related task and transfer the knowledge~\cite{LeeEtAl2021,WuEtAl2022};
(3) self supervision, where models learns from pseudo or weak labels~\cite{WuEtAl2021,ShuaiEtAl2022};
(4) few-shot learning, i.e., (meta-)learning from many related tasks with the aim of improving the performance on the problem of interest~\cite{SunEtAl2021,SunXZEtAl2021,SankarEtAl2021};
(5) exploiting prior knowledge manually encoded, for example external side information and Knowledge Graphs (KGs)~\cite{WangEtAl2019,AnelliEtAl2021}.
However, except for hand-coded knowledge, the above approaches still require a considerable amount of initial data or access to a different, but similar domain, where plenty of data is available.
On the other side, knowledge bases are application dependent, require access to expert knowledge, and are not always available. 

In this paper, we \textbf{contribute} a novel recommender approach able to operate in small data scenarios: our model does not require large volumes of initial data and is not application dependent.
We use a heterogeneous graph, where vertices denote entities, e.g., users and different types of content, and edges represent interactions between users and content, e.g., a user posting a message on a social media.
Then we use Personalized PageRank (PPR) to recommend items.
Note that, edges represent any interaction with users and content, not only interaction with recommendable items. 
We assign weights to edges in the graph, which represent the strength of the relationship between users and content.
In previous work within RSs~\cite{Lee2013,Xiang2010,Yao2013}, such weights are usually pre-defined depending on the application.
We do not make any assumption on the values of such weights and optimize them with a genetic algorithm~\cite{genetic_algorithm}.
To the best of our knowledge, heuristic algorithms have never been applied to learn edge weights in the context of RSs.
Our approach is evaluated on two real-world use cases: (1) an emergent educational social network, where there are few user interactions due to the initial stage of the platform, but a large number of items; (2) an insurance e-commerce platform, where there are many users but few user interactions, because users do not interact often with insurance products, and few items by nature of the insurance domain.
Experimental results are promising, showing up to a $7\%$ improvement over state-of-the-art baselines.

\section{Related Work}

Recommendation with small data has been tackled heuristically, i.e., by recommending items based on a set of specific rules~\cite{InozemtsevaEtAl2014}.
Such rules have to be designed for each use case, making these models application dependent.
Hybrid RSs have also been proposed for small data, for instance by merging Content Based (CB) and association rules~\cite{KaminskasEtAl2015,KaminskasEtAl2017}. 
Note that the datasets in~\cite{KaminskasEtAl2015,KaminskasEtAl2017} are not publicly available.
Item-to-item recommendation is addressed in~\cite{SchnabelAndBennett2020} with a CF approach as a counterfactual problem, where a small collection of explicit user preferences is used to improve propensity estimation.
We cannot use this in our work because: (1) our task is not item-to-item recommendations; (2) we do not have access to explicit user preferences; (3) a large dataset (MovieLens 25M~\cite{HarperEtAl2016}) is still needed to estimate propensity (the small annotated dataset is only used to debias the propensity estimate).
In~\cite{StrickrothAndPinkwart2012}, a hybrid user-based model combines CF, rule-based recommendation, and the top popular recommender with domain-specific and contextual information in the area of a small online educational community. The dataset is not publicly available and the approach is domain-dependent, hence not applicable to our work.
Finally, conversion rate prediction for small-scale recommendation is used in~\cite{PanEtAl2022}, with an ensemble of deep neural networks that are trained and evaluated on a non-public dataset of millions of users, impressions, and clicks.
Our small data scenario does not include enough data to train this ensemble model. 

Solutions for cold start cases (where users or items have few or no interactions) are hybrid combinations of CF, CB, demographic and contextual information~\cite{RaziperchikolaeiEtAl2021,BarkanEtAl2019,10.1145/3397271.3401060}, or ML methods such as data augmentation~\cite{XiaEtAl2022}, transfer learning~\cite{AggarwalEtAl2019,WuEtAl2022}, etc.~(see \S\ref{sec:intro}).
Data augmentation is used in~\cite{LeeEtAl2009}, where a CF model creates synthetic user ratings and is then combined with a CB model.
We cannot use this in our task because we do not have explicit ratings (we use any user interaction as implicit feedback).
In~\cite{WuEtAl2021}, self-supervision and data augmentation are combined on the user-item graph, and in~\cite{ShuaiEtAl2022}, self-supervision on the user-item graph is enhanced with features extracted from user reviews.
Few shot learning and meta-learning have also been used. 
In~\cite{SankarEtAl2021}, a neural recommender is trained over head items with frequent interactions, and this meta-knowledge is transferred to learn prototypes for long-tail items. In~\cite{SunXZEtAl2021}, recommendations for cold users are generated with a meta-learner that accounts for interest drift and geographical preferences.
In~\cite{AnelliEtAl2021}, knowledge bases (KG) are used to enrich feature representations, and in~\cite{WangEtAl2019} a neural attention mechanism learns the high order relation in the user-item graph and the KG.

%

All above approaches~\cite{XiaEtAl2022,AggarwalEtAl2019,WuEtAl2022,WuEtAl2021,ShuaiEtAl2022,SankarEtAl2021,SunXZEtAl2021,AnelliEtAl2021,WangEtAl2019} are evaluated on popular publicly available datasets, e.g., MovieLens~\cite{HarperEtAl2016}, Yelp, Amazon, CiteULike~\cite{VolkovsEtAl2017}, Weeplaces, etc.
These datasets are much larger than those in our case (see Tables~\ref{tab:number-interactions-educational social network} and~\ref{tab:number-interactions-insurance}) and allow using self-supervision, few-shot learning, attention mechanisms, and other neural models that we cannot use due to the extremely low amount of data.
Transfer learning and domain adaptation require large amounts of training data from a similar task or a related domain, which are not (publicly) available for our use cases.

Lastly, graph-based RSs can be robust as they enable information to propagate through vertices, unlike matrix completion which is affected by data sparsity~\cite{ShoujinWangEtAl2021}. 
This motivates recent approaches using GNN~\cite{WangEtAl2021,Salha-GalvanEtAl2021,ShuaiEtAl2022,WuEtAl2021,XiaEtAl2022}.
However, these are not applicable to small data problems because there are not enough samples to train GNN models.
PathRank~\cite{Lee2013} uses a heterogeneous user-item graph with additional vertices that are attributes of items, e.g., movie genre, director, etc.
Recommendations are generated with a random walk similar to PPR, but constraints are used to ensure that the random walks follow certain predefined paths. These are application dependent.
In~\cite{Yu2014}, the user-item graph is extended with item attributes, and meta-paths are defined to determine how two entities in the graph (vertices of different types) are connected (this encodes entity similarity).
A preference diffusion score is defined for specific meta-paths, based on user implicit feedback and co-occurrences of entities, and used to recommend items.
Unlike~\cite{Lee2013,Yu2014}, we build the heterogeneous graph from all user interactions, not only interactions with items. 
We also do not include item attributes in the graph and we do not use predefined paths or meta-paths. We assume any possible path and optimize edge weights with a genetic algorithm.

Injected Preference Fusion (IPF)~\cite{Xiang2010} extends PPR with a session-based temporal graph (STG) that includes both long- and short-term user preferences.
STG is a bipartite graph where users, items, and sessions are vertices. 
Non-negative weights are associated with edges, which control the balance between long- and short-term preferences.
%
Multi-Layer Context Graph (MLCG)~\cite{Yao2013} is a three-layer graph, where each layer represents a different type of context: (1) user context, e.g., gender and age; (2) item context, e.g., similarity between items; and (3) decision context, e.g., location and time.
Different weights are associated with intra- and inter-layer edges, defined as functions of vertice co-occurrence.
Unlike~\cite{Xiang2010,Yao2013}, (1) we represent in the graph all user interactions (not only those with recommendable items); (2) we do not consider temporal, contextual, or demographic features, which may not be available; (3) we do not use predefined weights, but we optimize them with a genetic algorithm.

\section{Approach}

Usually, graph-based recommendation consists of $2$ steps: (1) building the graph structure (\S\ref{subsec:build_graph}), and (2) recommending items (\S\ref{sec:generating-recommendations}).
We introduce an intermediate step between (1) and (2), where we optimize weights associated to different edge types (\S\ref{subsec:edge_weights}).

\subsection{Heterogeneous graph representation of user interactions}
\label{subsec:build_graph}

Let us consider a set of users who interact with content of various types, for example posts and comments in social networks or items and services in an online store.
We represent users and content (vertices), and their relationships (edges) with a heterogeneous graph~\cite{SunAndHan2012}. 
Vertices belong to different types, e.g., users, items, posts, etc.
Edges have different types depending on the action that they represent, e.g., the edge with the type ``like'' can connect a user with a post.
Moreover, edges have a direction because some actions are not symmetric, e.g., a user can follow another user, but not be followed by the same user.
Note that all types of user interactions are included in the graph, i.e., also interactions with objects that are not recommendable items, for example, a user creating a post or reporting a claim.

A heterogeneous graph, or heterogeneous information
network, is a type of directed graph $G = (\mathcal{V}, \mathcal{E}, \mathcal{A}, \mathcal{R})$, where vertices and edges represent different types of entities and relationships among them.
Each vertex $v\in \mathcal{V}$ and edge $e\in \mathcal{E}$ is associated to its type through a mapping function $\tau \colon \mathcal{V} \to \mathcal{A}$ and $\phi \colon \mathcal{E} \to \mathcal{R}$ respectively. 
$\mathcal{A}$ is the set of vertex types, or tags, e.g., users or various type of content. 
$\mathcal{R}$ is the set of edge types, e.g., a user liking a post.

We denote edges as $e = (i,j)$, where $i$ and $j\in \mathcal{V}$ and $i\neq j$.
Edge types are mapped to positive weights representing the strength of the relationship between two vertices. 
Formally, given an edge $(i,j)$, we define a weight function $W \colon \mathcal{R} \to \mathbb{R}^{+}$ such that $W(\phi((i,j))) = w_{i,j}$, with the constraint that $w_{i,j} >0$ (\S\ref{subsec:edge_weights} explains how to compute optimal weights $w_{i,j}$).

Since $G$ is a directed graph, $\phi((i,j)) \neq \phi((j,i))$, i.e., $\mathcal{R}$ contains distinct types for ingoing and outgoing edges.
Therefore, each user interaction is represented as two weighted edges\footnote{Except for non-symmetric actions, e.g., following a user.}, $w_{i, j}$ from the user to another vertex and $w_{j,i}$ from that vertex to the user vertex, e.g., a user liking a post and a post being liked by a user. 
The weights for those outgoing and ingoing edges might differ. 
Edges can also exist between two content vertices when two entities are related (for example a comment that was created under a post). 
In case of multiple interactions between $2$ vertices, e.g., a user can both create a post and like it, we define a different type of edge, i.e., a new value in $\mathcal{R}$, which represents the two actions.
The weight of such edge corresponds to the sum of the weights of each individual type, e.g., the creation and liking of a post.
In practice, this happens only for a few actions and does not affect the size of $\mathcal{R}$ significantly.

\begin{figure}[tb]
    \centering
    \includegraphics[width=0.75\textwidth]{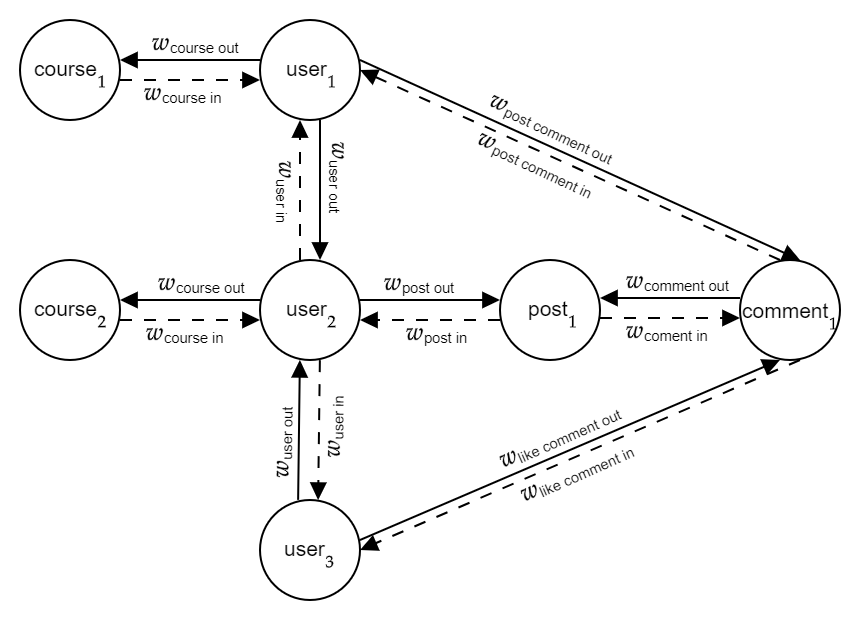}
    \caption{Example of the heterogeneous graph in a social network.}
    \label{fig:graph example}
\end{figure}
Figure~\ref{fig:graph example} illustrates an example of a heterogeneous graph in an educational social network, where user 1 follows course 1 (note that course means university course in an educational setting), user 2 follows course 2, and user 1 and 3 follow user 2. Furthermore, user 2 has created post 1, user 1 has created comment 1 under post 1 and user 3 has liked comment 1.

\subsection{Generating recommendations using random walks} 
\label{sec:generating-recommendations}

Given a user, the recommendation task consists in ranking vertices from the heterogeneous graph. 
To do this, we use PPR~\cite{Page1999}.
Starting at a source vertex $s$, the PPR value of a target vertex, $t$, is defined as the probability that an $\alpha$-discounted random walk from vertex $s$ terminates at $t$.  
An $\alpha$-discounted random walk represents a random walker that, at each step, either terminates at the current vertex with probability $\alpha$, or moves to a random out-neighbor with probability $1-\alpha$.
Formally, let $G$ be a graph of $n$ vertices, let $O(i)$ denote the set of end vertices of the outgoing edges of vertex $i$, and let the edge $(i,j)$ be weighted by $w_{ij}>0$. The steady-state distribution of an $\alpha$-discounted random walk in $G$, starting from vertex $s$, is defined as follows:

\begin{equation}
    \pi = (1-\alpha)P^T\pi+\alpha e_s \quad \text{where} \quad P = (p_{i,j})_{i,j \in V} = \frac{w_{i,j}}{\sum_{k \in O(i)}w_{i,k}} \cdot \mathbbm{1}_{\{j \in O(i)\}}
    \label{equ:ppr}
\end{equation}

$\alpha \in (0,1)$, $P$ is the transition matrix, $e_s$ is a one-hot vector of length $n$ with $e_s(s)=1$, and $\mathbbm{1}$ is the indicator function, equal to one when $j \in O(i)$.
Equation~\eqref{equ:ppr} is a linear system that can be solved using the power-iteration method~\cite{Page1999}.

Solving Equation~\eqref{equ:ppr} returns a $\pi$ for each user containing the PPR values (i.e., the ranks) of all the content vertices with respect to that user.
A recommendation list is then generated by either ordering the content vertices by their ranks and selecting the top-k, 
or by selecting the most similar neighbors by their ranks, then ordering the content by the neighbors' interaction frequency with the content. 
We implement both methods (see \S\ref{subsec:baselines}). 




\subsection{Optimizing edge weights using genetic algorithm}
\label{subsec:edge_weights}

Next, we explain how to compute the weight function $W$, which assigns the optimal weights for outgoing and ingoing edges of each interaction type.
In our data (which is presented in \S\ref{subsec:dataset}), the number of interaction types was $11$ and $9$, which required to optimize respectively $22$ and $18$ parameters (see Table~\ref{tab:interaction-weights}). 
With such a large search space, using grid search and similar methods would be very inefficient. Instead, we use a heuristic algorithm to find the optimal weights. Heuristic methods can be used to solve optimization problems that are not well suited for standard optimization algorithms, for instance, if the objective function is discontinuous or non-differentiable.
In particular, we use a genetic algorithm~\cite{genetic_algorithm} as our optimization algorithm, as it is a widely known and used algorithm, which is relatively straightforward to get started with and has been shown to serve as a strong baseline in many use cases~\cite{Odili2018}.
The algorithm in~\cite{genetic_algorithm} consists of $5$ components, which in our case are specified as follows: \textbf{1.~Initial population:} A population consisting of a set of gene vectors is initialized. In our case, each gene vector is a vector of weights with size $|\mathcal{R}|$, and each gene is uniformly initialized from a predefined range.
\textbf{2.~Fitness function:} Each of the initialized gene vectors is evaluated. In our case, recommendations are generated with PPR as described in \S\ref{sec:generating-recommendations} where the graph is weighted by the genes. The fitness function can be any evaluation measure that evaluates the quality of the ranked list of recommendations, e.g., normalized Discounted Cumulative Gain (nDCG), Mean Average Precision (MAP), etc.
\textbf{3.~Selection:} Based on the fitness score, the best gene vectors are selected to be parents for the next population.
\textbf{4.~Crossover:} Pairs of parents are mated with a uniform crossover type, i.e., offspring vectors are created where each gene in the vector is selected uniformly at random from one of the two mating parents.
\textbf{5.~Mutation:} Each gene in an offspring vector has a probability of being mutated, meaning that the value is modified by a small fraction. This is to maintain diversity and prevent local optima.
Finally, new offspring vectors are added to the population, and step $2$ to $5$ are repeated until the best solution converges.

\section{Experiments}

Next we describe the experimental evaluation: the use cases and datasets in \S\ref{subsec:dataset}; training and evaluation in \S\ref{sec:evaluation}; baselines and hyperparameters in \S\ref{subsec:baselines}; and results in \S\ref{subsec:results}. The code is publicly available\footnote{\url{https://github.com/simonebbruun/genetically_optimized_graph_RS}}.

\subsection{Use Cases and Datasets}
\label{subsec:dataset}

To evaluate our model, we need a dataset that satisfies the following criteria: (1) interaction scarcity; (2) different types of actions which might not be directly associated with items.
To the best of our knowledge, most publicly available datasets include only clicks and/or purchases or ratings, so they do not satisfy the second criterion.
We use two real-world datasets described next. 
The \textbf{educational social network} dataset was collected from a social platform for students between March 17, 2020 to April 6, 2022. 
We make this dataset public available\footnote{\url{https://github.com/carmignanivittorio/ai_denmark_data}}.
The users can, among others, follow courses from different universities, create and rate learning resources, and create, comment and like posts. 
The content vertices are: courses, universities, resources, posts, and comments and the goal is to recommend courses.  
The platform is maintained by a SME in the very early stage of growth and the dataset from it contains $5088$ interactions, made by $878$ different users with $1605$ different content objects resulting in a data sparsity of $0.996$. 
Dataset statistics are reported in Table~\ref{tab:number-interactions-educational social network}.


The second dataset is an \textbf{insurance} dataset~\cite{BruunEtAl2022} collected from an insurance vendor between October 1, 2018 to September 30, 2020\footnote{\url{https://github.com/simonebbruun/cross-sessions_RS}}. 
The content vertices are items and services within a specified section of the insurance website being either e-commerce, claims reporting, information or personal account.
Items are insurance products (e.g., car insurance) and additional coverages of insurance products (e.g., roadside assistance). 
Services can, among others, be specification of ``employment'' (required if you have an accident insurance), and information about ``the insurance process when moving home''.
User interactions are purchases and clicks on the insurance website.
The goal is to recommend items.
%
The dataset contains $432249$ interactions, made by $53569$ different users with $55$ different item and service objects resulting in a data sparsity of $0.853$. 
Dataset statistics are reported in Table~\ref{tab:number-interactions-insurance}.

\begin{table}[tb]
    \caption{Dataset statistics: educational social network.}
    \label{tab:number-interactions-educational social network}
\resizebox{\textwidth}{!}{%
\begin{tabular}{|c|cc|cc|cc|cc|c|}
\hline
\multirow{2}{*}{\begin{tabular}[c]{@{}c@{}}Type of\\ interaction\end{tabular}} & \multicolumn{2}{c|}{Follow} & \multicolumn{2}{c|}{Post} & \multicolumn{2}{c|}{Comment} & \multicolumn{2}{c|}{Source} & \multirow{2}{*}{Join University} \\ \cline{2-9}
 & \multicolumn{1}{c|}{Course} & User & \multicolumn{1}{c|}{Create} & Like & \multicolumn{1}{c|}{Create} & Like & \multicolumn{1}{c|}{Create} & Rate &  \\ \hline
Training & \multicolumn{1}{c|}{\begin{tabular}[c]{@{}c@{}}1578\\ (28.12\%)\end{tabular}} & \multirow{3}{*}{\begin{tabular}[c]{@{}c@{}}842\\ (15\%)\end{tabular}} & \multicolumn{1}{c|}{\multirow{3}{*}{\begin{tabular}[c]{@{}c@{}}92\\ (1.64\%)\end{tabular}}} & \multirow{3}{*}{\begin{tabular}[c]{@{}c@{}}339\\ (6.04\%)\end{tabular}} & \multicolumn{1}{c|}{\multirow{3}{*}{\begin{tabular}[c]{@{}c@{}}116\\ (2.07\%)\end{tabular}}} & \multirow{3}{*}{\begin{tabular}[c]{@{}c@{}}96\\ (1.71\%)\end{tabular}} & \multicolumn{1}{c|}{\multirow{3}{*}{\begin{tabular}[c]{@{}c@{}}75\\ (1.34\%)\end{tabular}}} & \multirow{3}{*}{\begin{tabular}[c]{@{}c@{}}113\\ (2.01\%)\end{tabular}} & \multirow{3}{*}{\begin{tabular}[c]{@{}c@{}}1400\\ (24.95\%)\end{tabular}} \\ \cline{1-2}
Validation & \multicolumn{1}{c|}{\begin{tabular}[c]{@{}c@{}}415\\ (7.39\%)\end{tabular}} &  & \multicolumn{1}{c|}{} &  & \multicolumn{1}{c|}{} &  & \multicolumn{1}{c|}{} &  &  \\ \cline{1-2}
Test & \multicolumn{1}{c|}{\begin{tabular}[c]{@{}c@{}}546\\ (9.73\%)\end{tabular}} &  & \multicolumn{1}{c|}{} &  & \multicolumn{1}{c|}{} &  & \multicolumn{1}{c|}{} &  &  \\ \hline
\end{tabular}%
}
\end{table}

\begin{table}[tb]
    \caption{Dataset statistics: insurance dataset.}
    \label{tab:number-interactions-insurance}
\resizebox{\textwidth}{!}{%
\begin{tabular}{|c|c|cc|cc|cc|cc|}
\hline
\multirow{2}{*}{\begin{tabular}[c]{@{}c@{}}Type of\\ interaction\end{tabular}} & \multirow{2}{*}{Purchase items} & \multicolumn{2}{c|}{E-commerce} & \multicolumn{2}{c|}{Personal account} & \multicolumn{2}{c|}{Claims reporting} & \multicolumn{2}{c|}{Information} \\ \cline{3-10} 
 &  & \multicolumn{1}{c|}{Items} & Services & \multicolumn{1}{c|}{Items} & Services & \multicolumn{1}{c|}{Items} & Services & \multicolumn{1}{c|}{Items} & Services \\ \hline
Training & \begin{tabular}[c]{@{}c@{}}4853\\ (13.65\%)\end{tabular} & \multicolumn{1}{c|}{\multirow{3}{*}{\begin{tabular}[c]{@{}c@{}}6897\\ (19.4\%)\end{tabular}}} & \multirow{3}{*}{\begin{tabular}[c]{@{}c@{}}1775\\ (4.99\%)\end{tabular}} & \multicolumn{1}{c|}{\multirow{3}{*}{\begin{tabular}[c]{@{}c@{}}287\\ (0.81\%)\end{tabular}}} & \multirow{3}{*}{\begin{tabular}[c]{@{}c@{}}17050\\ (47.96\%)\end{tabular}} & \multicolumn{1}{c|}{\multirow{3}{*}{\begin{tabular}[c]{@{}c@{}}154\\ (0.43\%)\end{tabular}}} & \multirow{3}{*}{\begin{tabular}[c]{@{}c@{}}6\\ (0.02\%)\end{tabular}} & \multicolumn{1}{c|}{\multirow{3}{*}{\begin{tabular}[c]{@{}c@{}}1129\\ (3.18\%)\end{tabular}}} & \multirow{3}{*}{\begin{tabular}[c]{@{}c@{}}2118\\ (5.96\%)\end{tabular}} \\ \cline{1-2}
Validation & \begin{tabular}[c]{@{}c@{}}601\\ (1.69\%)\end{tabular} & \multicolumn{1}{c|}{} &  & \multicolumn{1}{c|}{} &  & \multicolumn{1}{c|}{} &  & \multicolumn{1}{c|}{} &  \\ \cline{1-2}
Test & \begin{tabular}[c]{@{}c@{}}680\\ (1.91\%)\end{tabular} & \multicolumn{1}{c|}{} &  & \multicolumn{1}{c|}{} &  & \multicolumn{1}{c|}{} &  & \multicolumn{1}{c|}{} &  \\ \hline
\end{tabular}%
}
\end{table}

\subsection{Evaluation Procedure} 
\label{sec:evaluation}
We split the datasets into training and test set as follows.
As test set for the educational social network dataset, we use the last course interaction (leave-one-out) for each user who has more than one course interaction. 
The remaining is used as training set.
All interactions occurring after the left-out course interaction in the test set are removed to prevent data leakage.
%
As test set for the insurance dataset, we use the latest 10\% of purchase events (can be one or more purchases made by the same user). The remaining interactions (occurring before the purchases in the test set) are used as training set.

For each user in the test set, the RS generates a ranked list of content vertices to be recommended.
For the educational social network dataset, 
courses that the user already follows are filtered out from the ranked list. 
For the insurance dataset, 
it is only possible for a user to buy an additional coverage if the user has the corresponding base insurance product, therefore we filter out additional coverages if this is not the case, as per~\cite{Aggarwal2016}. 

As evaluation measures, we use Hit Rate (HR) and Mean Reciprocal Rank (MRR). 
Since in most cases, we only have one true content object for each user in the test set (leave-one-out), MRR corresponds to MAP and is somehow proportional to nDCG (they differ in the discount factor). 
For the educational social network, we use standard cutoffs, i.e.,~$5$ and $10$. For the insurance dataset, we use a cutoff of $3$ because the total number of items is $16$, therefore with higher cut-offs all measures will reach high values, which will not inform on the actual quality of the RSs.

\subsection{Baselines, Implementation, and Hyperparameters}
\label{subsec:baselines}




The focus of this work is to improve the quality of recommendations on small data problems, such as the educational social network dataset. Therefore, we consider both simple collaborative filtering baselines that are robust on small datasets as well as state-of-the-art neural baselines: \textit{Most Popular} recommends the content with most interactions across users; \textit{UB-KNN} is a user-based nearest neighbor model that computes similarities between users, then ranks the content by the interaction frequency of the top-k neighbors. Similarity is defined as the cosine similarity between the binarized vectors of user interactions; \textit{SVD} is a latent factor model that factorizes the matrix of user interactions by singular value decomposition~\cite{CremonesiEtAl2010}; \textit{NeuMF} factorizes the matrix of user interactions and replaces the user-item inner product with a neural architecture~\cite{He2017}; \textit{NGCF} represents user interactions in a bipartite graph and uses a graph neural network to learn user and item embeddings~\cite{Wang2019}; \textit{Uniform Graph} is a graph-based model that ranks the vertices using PPR~\cite{Page1999} with all edge weights equal to $1$;
\textit{User Study Graph} is the same as uniform, but the weights are based on a recent user study conducted on the same educational social network~\cite{anonymous2020}.
Users assigned $2$ scores to each action type: the effort required to perform the action, and the value that the performed action brings to the user.
We normalized the scores and used effort scores for outgoing edges and value scores for ingoing edges.
The exact values are in our source code.
A similar user study is not available for the insurance domain.

%
%

All implementation is in \texttt{Python} \texttt{3.9}. 
%
Hyperparameters are tuned on a validation set, created from the training set in the same way as the test set (see \S\ref{sec:evaluation}). 
For the educational social network, optimal hyperparameters are the following: damping factor $\alpha = 0.3$; PPR best predictions are obtained by ranking vertices; $30$ latent factors for SVD; and number of neighbors $k=60$ for UB-KNN.
Optimal hyperparametrs in the insurance dataset are: damping factor $\alpha = 0.4$; PPR best predictions are obtained by ranking user vertices and select the closest $90$ users; $10$ latent factors for SVD; and number of neighbors $k=80$ for UB-KNN.


The genetic algorithm is implemented with \texttt{PyGAD}  \texttt{2.16.3} with MRR as fitness function and the following parameters: initial population: $10$; gene range: $[0.01,2]$, parents mating: $4$; genes to mutate: $10\%$; mutation range: $[-0.3,0.3]$.
We optimize the edge weights using the training set to build the graph and the validation set to evaluate the fitness function. 
The optimal weights are reported in Table~\ref{tab:interaction-weights}. In order to provide stability of the optimal weights, we report the average weights obtained by five runs of the genetic algorithm. 


\begin{table}[tb]
\centering
    \caption{Optimized interaction weights averaged over five runs of the genetic algorithm.}
    \label{tab:interaction-weights}
    \begin{subtable}[t]{.46\textwidth}
    \centering
\caption{Educational social network dataset.}
\label{tab:interaction-weights-educational social network}
\resizebox{\textwidth}{!}{
\begin{tabular}{|c|cc|}
\hline
Trained   for & \multicolumn{2}{c|}{Course follows} \\ \hline
Direction of edge & \multicolumn{1}{c|}{Out} & In \\ \hline
User follows user & \multicolumn{1}{c|}{0.95} & 0.86 \\ \hline
User follows course & \multicolumn{1}{c|}{0.73} & 1.88 \\ \hline
User creates post & \multicolumn{1}{c|}{1.11} & 1.09 \\ \hline
User creates resource & \multicolumn{1}{c|}{1.27} & 0.55 \\ \hline
User creates comment & \multicolumn{1}{c|}{0.91} & 1.03 \\ \hline
User likes post & \multicolumn{1}{c|}{1.27} & 0.61 \\ \hline
User likes comment & \multicolumn{1}{c|}{0.42} & 0.99 \\ \hline
User rates resource & \multicolumn{1}{c|}{0.77} & 0.84 \\ \hline
Comment under post & \multicolumn{1}{c|}{0.91} & 1.17 \\ \hline
User joins university & \multicolumn{1}{c|}{0.28} & 1.06 \\ \hline
\end{tabular}
}
    \end{subtable}%
    \hfill
    \begin{subtable}[t]{.51\textwidth}
    \centering
\caption{Insurance dataset.}
\label{tab:interaction-weights-insurance}
\resizebox{\textwidth}{!}{
\begin{tabular}{|c|cc|}
\hline
Trained for & \multicolumn{2}{c|}{Purchase items} \\ \hline
Direction of edge & \multicolumn{1}{c|}{Out} & In \\ \hline
Purchase items & \multicolumn{1}{c|}{0.24} & 1.05 \\ \hline
E-commerce items & \multicolumn{1}{c|}{0.64} & 1.49 \\ \hline
E-commerce services & \multicolumn{1}{c|}{1.21} & 1.18 \\ \hline
Personal account items & \multicolumn{1}{c|}{1.06} & 0.36 \\ \hline
Personal account services & \multicolumn{1}{c|}{0.92} & 0.66 \\ \hline
Claims reporting items & \multicolumn{1}{c|}{0.93} & 0.58 \\ \hline
Claims reporting services & \multicolumn{1}{c|}{0.90} & 1.32 \\ \hline
Information items & \multicolumn{1}{c|}{1.60} & 0.79 \\ \hline
Information services & \multicolumn{1}{c|}{0.64} & 0.51 \\ \hline
\end{tabular}
}
    \end{subtable} 
\end{table}

\subsection{Results}
\label{subsec:results}

Table~\ref{tab:results} reports experimental results.
%
On both datasets, UB-KNN outperforms SVD and NeuMF, and the best-performing baseline is the uniform graph-based model on the educational social network dataset and the NGCF model on the insurance dataset. 
This corroborates previous findings, showing that graph-based RSs are more robust than matrix factorization when data is sparse~\cite{ShoujinWangEtAl2021} and neural models need a considerable amount of data to perform well.
Graph-based methods account for indirect connectivity among content vertices and users, thus outperforming also UB-KNN, which defines similar users on subsets of commonly interacted items.
Our genetically optimized graph-based model outperforms all baseline models on the educational social network dataset and obtains competing results with the NGCF model on the insurance dataset, showing that the genetic algorithm can successfully find the best weights, which results in improved effectiveness. In order to account for randomness of the genetic algorithm, we run the optimization of weights five times and report the mean and standard deviation of the results in Table~\ref{tab:results}. The standard deviation is lowest on the insurance dataset, but even on the very small educational dataset, the standard deviation is relatively low, so for different initializations, the algorithm tends to converge toward similar results.
Moreover, we tried a version of our model where we let the graph be an undirected graph, meaning that for each edge type the weight for the ingoing and the outgoing edge is the same. The results show that the directed graph outperforms its undirected version.
For the educational social network, weights based on the user study result in worse performance, even lower than UB-KNN.
Overall, scores are higher on the insurance data.
This might happen because: (1) data from the educational social network is sparser than insurance data (see \S\ref{subsec:dataset}); (2) the insurance data has a considerably larger training set; (3) there are fewer items to recommend in the insurance domain ($16$ vs.~$388$).

\begin{figure}[tb]
     \hspace*{\fill}%
    \begin{subfigure}[t]{0.49\textwidth}
        \includegraphics[width=\columnwidth]{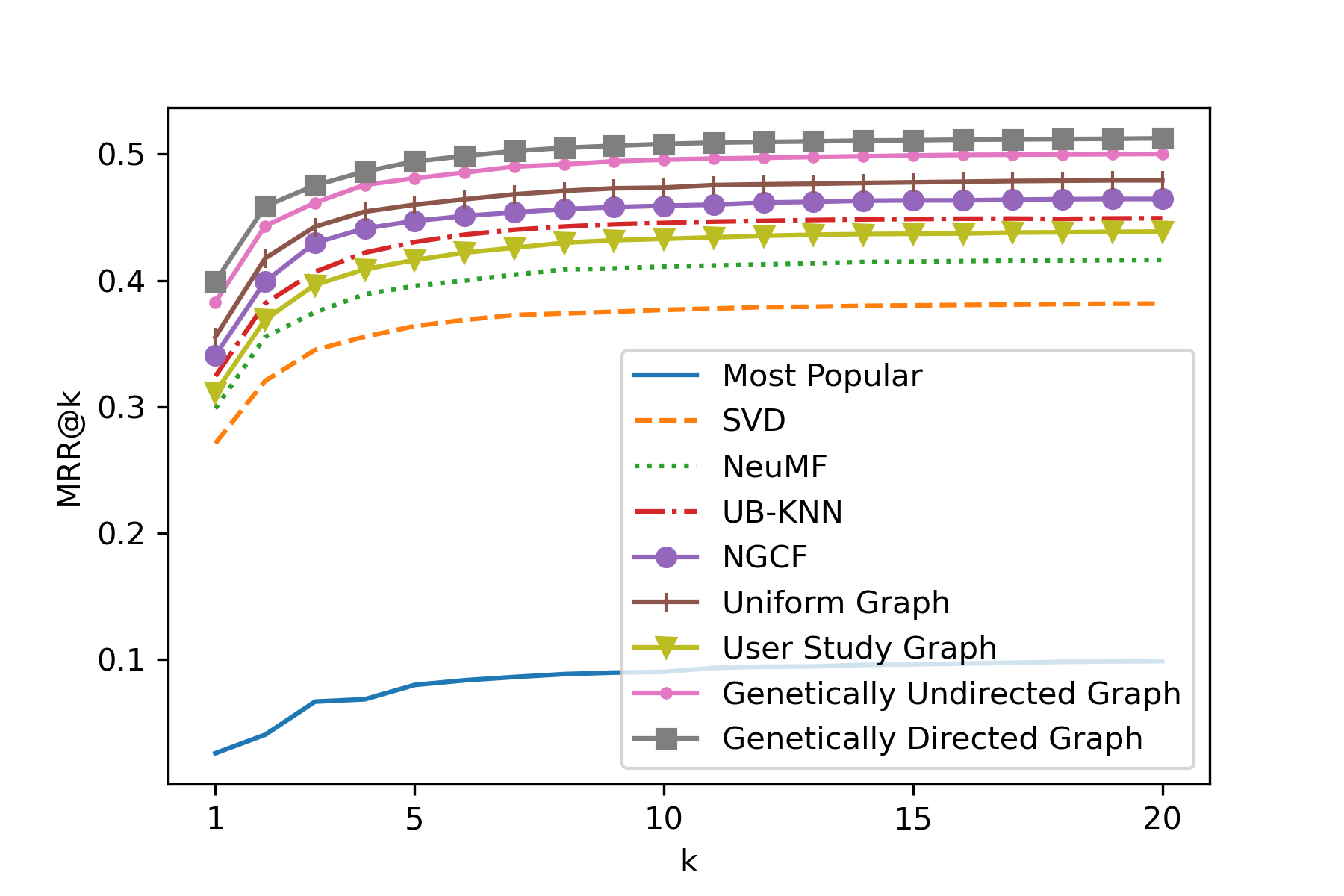}
        \caption{Educational social network dataset.}
    \end{subfigure}
    \hfill%
    \begin{subfigure}[t]{0.49\textwidth}
        \includegraphics[width=\columnwidth]{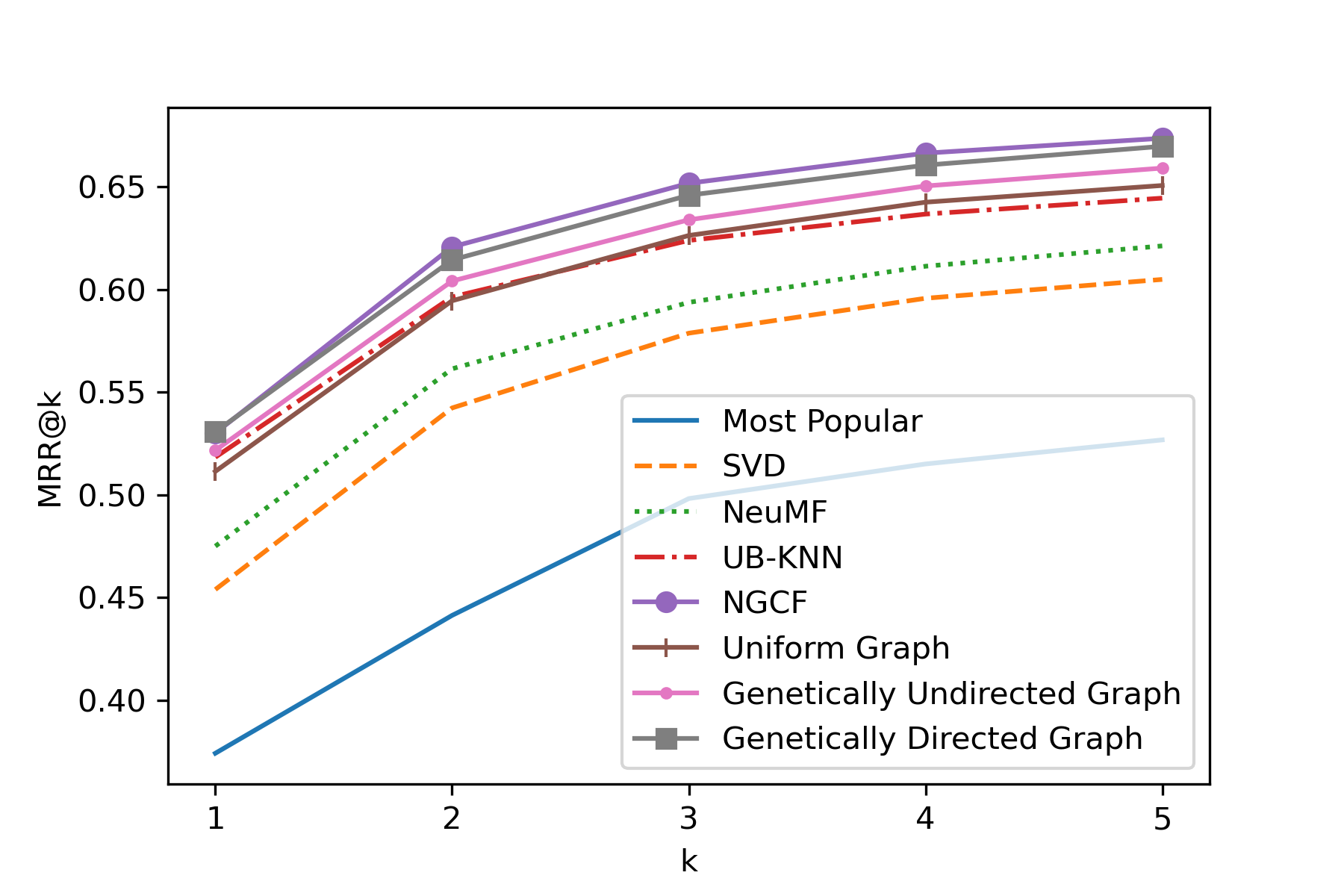}
        \caption{Insurance dataset.}
    \end{subfigure}
    \hspace*{\fill}%
    \caption{MRR@k for varying choices of the cutoff threshold $k$.}
     \label{fig:varying cutoffs}
\end{figure}

Figure~\ref{fig:varying cutoffs} shows MRR at varying cutoffs $k$. We have similar results for HR, which are omitted due to space limitations. It appears that the results are consistent for varying thresholds. 
Only on the insurance dataset, we see that the UB-KNN is slightly better than the uniform graph-based model for smaller thresholds. 

Inspecting the optimal weights in Table~\ref{tab:interaction-weights}, we see that for the educational social network all the interaction types associated with courses (following course, creating resource, creating comment, creating and liking posts) are highly weighted. 
This is reasonable since courses are the recommended items.
Moreover, a higher weight is assigned when a user follows a user compared to when a user is followed by a user. 
This reasonably suggests that users are interested in courses attended by the users they follow, rather than the courses of their followers.
For the insurance dataset, we observe that the greatest weights are given when a user clicks on items in the information and personal account section, when items are purchased by a user, and when items and services are clicked in the e-commerce section, which are all closely related to the process of purchasing items.

\begin{table}[tb]
\caption{Performance results ($^\dag$mean/std). All results marked with * are significantly different with a confidence level of $0.05$ from the genetically optimized graph (ANOVA~\cite{kutner2005} is used for MRR@k and McNemar's test~\cite{Dietterich1998} is used for HR@k). The best results are in bold.}
\label{tab:results}
\resizebox{\textwidth}{!}{
\begin{tabular}{|c|cccc|cc|}
\hline
Dataset & \multicolumn{4}{c|}{Educational social   network} & \multicolumn{2}{c|}{Insurance} \\ \hline
Measure & \multicolumn{1}{c|}{MRR@5} & \multicolumn{1}{c|}{MRR@10} & \multicolumn{1}{c|}{HR@5} & HR@10 & \multicolumn{1}{c|}{MRR@3} & HR@3 \\ \hline
Most Popular & \multicolumn{1}{c|}{0.0797*} & \multicolumn{1}{c|}{0.0901*} & \multicolumn{1}{c|}{0.1978*} & 0.2729* & \multicolumn{1}{c|}{0.4982*} & 0.6791* \\ \hline
SVD & \multicolumn{1}{c|}{0.3639*} & \multicolumn{1}{c|}{0.3767*} & \multicolumn{1}{c|}{0.5275*} & 0.6209* & \multicolumn{1}{c|}{0.5787*} & 0.7399* \\ \hline
NeuMF & \multicolumn{1}{c|}{0.3956*} & \multicolumn{1}{c|}{0.4110*} & \multicolumn{1}{c|}{0.5604*} & 0.6740* & \multicolumn{1}{c|}{0.5937*} & 0.7448* \\ \hline
UB-KNN & \multicolumn{1}{c|}{0.4304*} & \multicolumn{1}{c|}{0.4456*} & \multicolumn{1}{c|}{0.6172*} & 0.7271* & \multicolumn{1}{c|}{0.6238*} & 0.7569* \\ \hline
NGCF & \multicolumn{1}{c|}{0.4471} & \multicolumn{1}{c|}{0.4592} & \multicolumn{1}{c|}{0.6245*} & 0.7143* & \multicolumn{1}{c|}{\textbf{0.6517}} & \textbf{0.8043}* \\ \hline
Uniform Graph & \multicolumn{1}{c|}{0.4600} & \multicolumn{1}{c|}{0.4735} & \multicolumn{1}{c|}{0.6300*} & 0.7289* & \multicolumn{1}{c|}{0.6263*} & 0.7730* \\ \hline
User Study Graph & \multicolumn{1}{c|}{0.4162*} & \multicolumn{1}{c|}{0.4330*} & \multicolumn{1}{c|}{0.5952*} & 0.7179* & \multicolumn{1}{c|}{-} & - \\ \hline
Genetically Undirected Graph & \multicolumn{1}{c|}{0.4809} & \multicolumn{1}{c|}{0.4957} & \multicolumn{1}{c|}{0.6410} & 0.7509 & \multicolumn{1}{c|}{0.6339} & 0.7760* \\ \hline
\begin{tabular}[c]{@{}c@{}}Genetically\\ Directed Graph$^\dag$\end{tabular} & \multicolumn{1}{c|}{\begin{tabular}[c]{@{}c@{}}\textbf{0.4907}/\\ 0.0039\end{tabular}} & \multicolumn{1}{c|}{\begin{tabular}[c]{@{}c@{}}\textbf{0.5045}/\\ 0.0037\end{tabular}} & \multicolumn{1}{c|}{\begin{tabular}[c]{@{}c@{}}\textbf{0.6505}/\\ 0.0052\end{tabular}} & \begin{tabular}[c]{@{}c@{}}\textbf{0.7520}/\\ 0.0055\end{tabular} & \multicolumn{1}{c|}{\begin{tabular}[c]{@{}c@{}}0.6435/\\ 0.0029\end{tabular}} & \begin{tabular}[c]{@{}c@{}}0.7875/\\ 0.0044\end{tabular} \\ \hline
\end{tabular}
}
\end{table}

We further evaluate the performance of our models, when trained on smaller samples of the insurance dataset. We randomly sample $10\%$, $25\%$ and $50\%$ from the training data, which is then split into training and validation set as described in \S\ref{subsec:baselines}. In order to account for randomness of the genetic algorithm, we sample $5$ times for each sample size and report the mean and standard deviation of the results. We evaluate the models on the original test set for comparable results. The results are presented in Table~\ref{tab:results_sample}. While the genetically optimized graph-based model only partially outperforms the NGCF model on large sample sizes ($50\%$ and $100\%$) it outperforms all the baselines on small sample sizes ($10\%$ and $25\%$). This shows that our genetically optimized model is more robust for small data problems than a neural graph-based model. In addition, it is still able to compete with the neural graph-based model when larger datasets are available.

\begin{table}[tb]
\caption{Results on smaller samples of the insurance dataset. The notation is as in Table~\ref{tab:results}.}
\label{tab:results_sample}
\resizebox{\textwidth}{!}{
\begin{tabular}{|c|cc|cc|cc|cc|}
\hline
\begin{tabular}[c]{@{}c@{}}Percentage of\\ insurance dataset\end{tabular} & \multicolumn{2}{c|}{10\%} & \multicolumn{2}{c|}{25\%} & \multicolumn{2}{c|}{50\%} & \multicolumn{2}{c|}{100\%} \\ \hline
Measure & \multicolumn{1}{c|}{MRR@3} & HR@3 & \multicolumn{1}{c|}{MRR@3} & HR@3 & \multicolumn{1}{c|}{MRR@3} & HR@3 & \multicolumn{1}{c|}{MRR@3} & HR@3 \\ \hline
Most popular & \multicolumn{1}{c|}{0.5003*} & 0.6856* & \multicolumn{1}{c|}{0.4981*} & 0.6789* & \multicolumn{1}{c|}{0.5003*} & 0.6856* & \multicolumn{1}{c|}{0.4982*} & 0.6791* \\ \hline
SVD & \multicolumn{1}{c|}{0.5785*} & 0.7336* & \multicolumn{1}{c|}{0.5794*} & 0.7395* & \multicolumn{1}{c|}{0.5758*} & 0.7379* & \multicolumn{1}{c|}{0.5787*} & 0.7399* \\ \hline
NeuMF & \multicolumn{1}{c|}{0.5792*} & 0.7326* & \multicolumn{1}{c|}{0.5759*} & 0.7291* & \multicolumn{1}{c|}{0.5849*} & 0.7466* & \multicolumn{1}{c|}{0.5937} & 0.7448 \\ \hline
UB-KNN & \multicolumn{1}{c|}{0.6183} & 0.7661* & \multicolumn{1}{c|}{0.6180*} & 0.7494* & \multicolumn{1}{c|}{0.6243} & 0.7524* & \multicolumn{1}{c|}{0.6238*} & 0.7569* \\ \hline
NGCF & \multicolumn{1}{c|}{0.5937*} & 0.7199* & \multicolumn{1}{c|}{0.6030*} & 0.7405* & \multicolumn{1}{c|}{\textbf{0.6397}} & 0.7894 & \multicolumn{1}{c|}{\textbf{0.6517}} & \textbf{0.8043*} \\ \hline
Uniform Graph & \multicolumn{1}{c|}{0.6196} & 0.7687* & \multicolumn{1}{c|}{0.6206*} & 0.7687* & \multicolumn{1}{c|}{0.6253} & 0.7746 & \multicolumn{1}{c|}{0.6263*} & 0.7730* \\ \hline
\begin{tabular}[c]{@{}c@{}}Genetically\\ Undirected Graph\end{tabular} & \multicolumn{1}{c|}{0.6238} & 0.7672* & \multicolumn{1}{c|}{0.6224} & 0.7601* & \multicolumn{1}{c|}{0.6286} & 0.7741 & \multicolumn{1}{c|}{0.6339} & 0.7760 \\ \hline
\begin{tabular}[c]{@{}c@{}}Genetically\\ Directed Graph$^\dag$\end{tabular} & \multicolumn{1}{c|}{\begin{tabular}[c]{@{}c@{}}\textbf{0.6267}/\\ 0.0052\end{tabular}} & \begin{tabular}[c]{@{}c@{}}\textbf{0.7784}/\\ 0.0084\end{tabular} & \multicolumn{1}{c|}{\begin{tabular}[c]{@{}c@{}}\textbf{0.6353}/\\ 0.0039\end{tabular}} & \begin{tabular}[c]{@{}c@{}}\textbf{0.7845}/\\ 0.0073\end{tabular} & \multicolumn{1}{c|}{\begin{tabular}[c]{@{}c@{}}\textbf{0.6397}/\\ 0.0045\end{tabular}} & {\begin{tabular}[c]{@{}c@{}}\textbf{0.7903}/\\ 0.0071\end{tabular}} & \multicolumn{1}{c|}{\begin{tabular}[c]{@{}c@{}}0.6435/\\ 0.0029\end{tabular}} & \begin{tabular}[c]{@{}c@{}}0.7875/\\ 0.0044\end{tabular} \\ \hline
\end{tabular}
}
\end{table}

\begin{figure}[tb]
    \centering
    \includegraphics[width=0.75\textwidth]{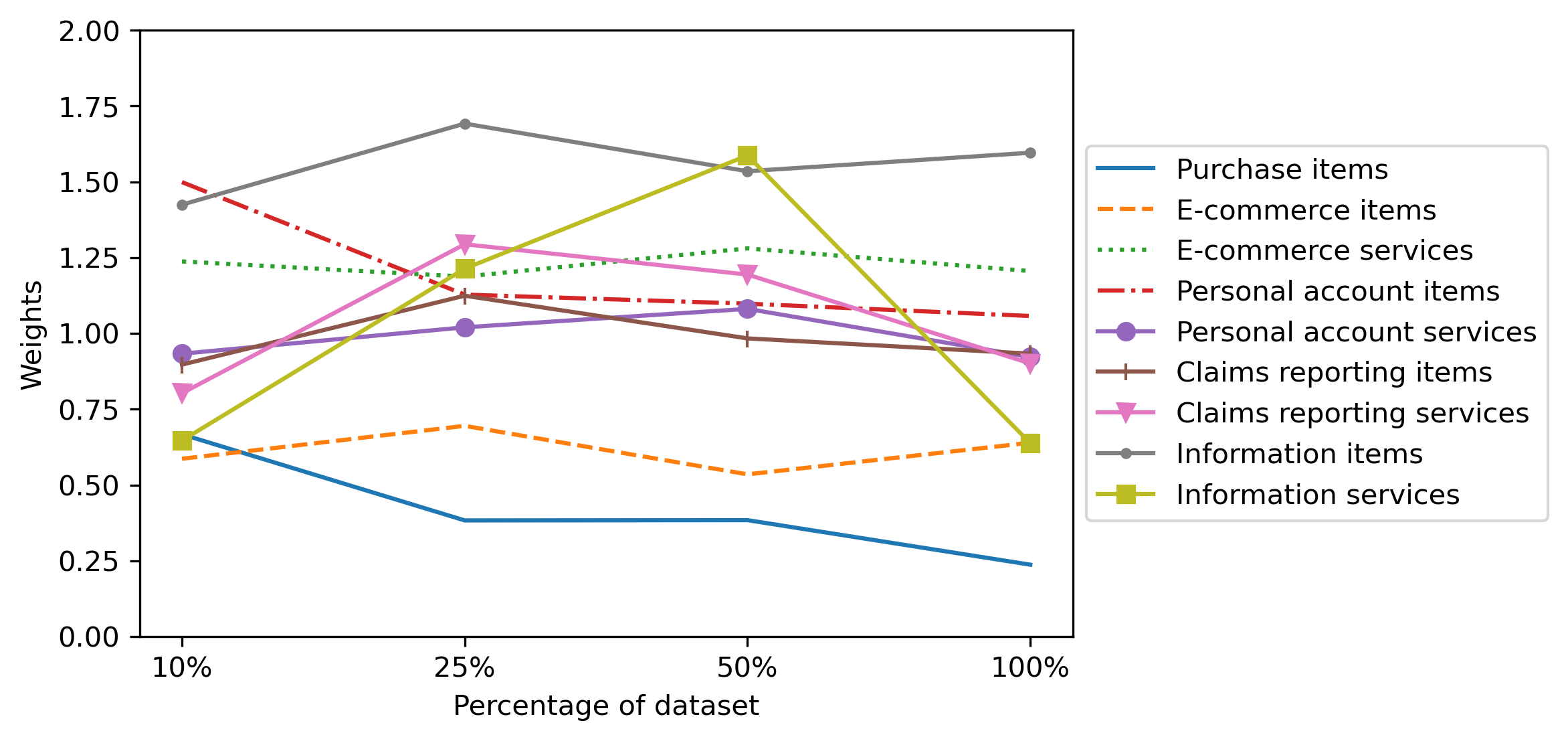}
    \caption{Plot of how the outgoing edge weights evolve for different sizes of the insurance dataset.}
    \label{fig:weights}
\end{figure}

In Figure~\ref{fig:weights} we inspect how the outgoing edge weights evolve when optimized on different sizes of the insurance dataset. We have similar results for the ingoing edge weights, which are omitted due to space limitation. It appears that the optimized weights only change a little when more data is added to the training set, and the relative importance of the interaction types remains stable across the different sizes of the dataset. Only the interaction type ``information services`` has large variations across the different dataset sizes, and in general, the biggest development of the weights happens when the dataset is increased from $10\%$ to $25\%$. It shows that once the genetic algorithm has found the optimal weights in offline mode, the weights can be held fixed while the RS is deployed online, and the weights only need to be retrained (offline) once in a while, reducing the need for a fast optimization algorithm.

\section{Conclusions and Future Work}
We have introduced a novel recommender approach able to cope with very low data scenarios. 
This is a highly relevant problem for SMEs that might not have access to large amounts of data.
We use a heterogeneous graph with users, content and their interactions to generate recommendations. 
We assign different weights to edges depending on the interaction type and use a genetic algorithm to find the optimal weights. 
Experimental results on two different use cases show that our model outperforms state-of-the-art baselines for two real-world small data scenarios. We make our code and datasets publicly available.

As future work we will consider possible extensions of the graph structure, for example, we can include contextual and demographic information as additional layers, similarly to what is done in~\cite{Yao2013}.
Moreover, we can account for the temporal dimension, by encoding the recency of the actions in the edge weight, as done in~\cite{Xiang2010}.
We will further experiment with the more recent particle swarm \cite{Kennedy1995} and ant colony optimization algorithms \cite{Dorigo2006} instead of the genetic algorithm to find the optimal weights.
Finally, we will investigate how to incorporate the edge weights into an explainability model, so that we can provide explanations to end users in principled ways as done in~\cite{Atanasova_Simonsen_Lioma_Augenstein_2022}.



\small{
\section*{Acknowledgements}
This paper was partially supported by the European Union’s Horizon 2020 research and innovation programme under the Marie Sklodowska-Curie grant agreement No. 893667 and the Industriens Fond, AI Denmark project.}

%
%
%
\bibliographystyle{splncs04}
\bibliography{mybibliography}

\begin{thebibliography}{10}
\providecommand{\url}[1]{\texttt{#1}}
\providecommand{\urlprefix}{URL }
\providecommand{\doi}[1]{https://doi.org/#1}

\bibitem{Aggarwal2016}
Aggarwal, C.C.: {Context-Sensitive Recommender Systems}, pp. 255--281. Springer
  International Publishing (2016). \doi{10.1007/978-3-319-29659-3\_8}

\bibitem{AggarwalEtAl2019}
Aggarwal, K., Yadav, P., Keerthi, S.S.: {Domain Adaptation in Display
  Advertising: an Application for Partner Cold-start}. In: Bogers, T., Said,
  A., Brusilovsky, P., Tikk, D. (eds.) Proceedings of the 13th {ACM} Conference
  on Recommender Systems, (RecSys 2019). pp. 178--186. {ACM} (2019),
  \url{https://doi.org/10.1145/3298689.3347004}

\bibitem{AnelliEtAl2021}
Anelli, V.W., Noia, T.D., Sciascio, E.D., Ferrara, A., Mancino, A.C.M.: {Sparse
  Feature Factorization for Recommender Systems with Knowledge Graphs}. In:
  Pamp{\'{\i}}n, H.J.C., Larson, M.A., Willemsen, M.C., Konstan, J.A., McAuley,
  J.J., Garcia{-}Gathright, J., Huurnink, B., Oldridge, E. (eds.) Proceedings
  of the 15th {ACM} Conference on Recommender Systems, (RecSys 2021). pp.
  154--165. {ACM} (2021), \url{https://doi.org/10.1145/3460231.3474243}

\bibitem{Atanasova_Simonsen_Lioma_Augenstein_2022}
Atanasova, P., Simonsen, J.G., Lioma, C., Augenstein, I.: Diagnostics-guided
  explanation generation. Proceedings of the AAAI Conference on Artificial
  Intelligence  \textbf{36}(10),  10445--10453 (Jun 2022).
  \doi{10.1609/aaai.v36i10.21287},
  \url{https://ojs.aaai.org/index.php/AAAI/article/view/21287}

\bibitem{BarkanEtAl2019}
Barkan, O., Koenigstein, N., Yogev, E., Katz, O.: {CB2CF: a Neural Multiview
  Content-to-Collaborative Filtering Model for Completely Cold Item
  Recommendations}. In: Bogers, T., Said, A., Brusilovsky, P., Tikk, D. (eds.)
  Proceedings of the 13th {ACM} Conference on Recommender Systems, (RecSys
  2019). pp. 228--236. {ACM} (2019),
  \url{https://doi.org/10.1145/3298689.3347038}

\bibitem{anonymous2020}
Biasini, M.: {Design and Implementation of Gamification in a Social e-Learning
  Platform for Increasing Learner Engagement}. Master's thesis, Danmarks
  Tekniske Universitet and Università degli Studi di Padova (2020)

\bibitem{BruunEtAl2022}
Bruun, S.B., Maistro, M., Lioma, C.: {Learning Recommendations from User
  Actions in the Item-poor Insurance Domain}. In: Golbeck, J., Harper, F.M.,
  Murdock, V., Ekstrand, M.D., Shapira, B., Basilico, J., Lundgaard, K.T.,
  Oldridge, E. (eds.) Proceedings of the 16th {ACM} Conference on Recommender
  Systems, (RecSys 2022). pp. 113--123. {ACM} (2022),
  \url{https://doi.org/10.1145/3523227.3546775}

\bibitem{CremonesiEtAl2010}
Cremonesi, P., Koren, Y., Turrin, R.: {Performance of Recommender Algorithms on
  Top-n Recommendation Tasks}. In: Amatriain, X., Torrens, M., Resnick, P.,
  Zanker, M. (eds.) Proceedings of the 4th {ACM} Conference on Recommender
  Systems, (RecSys 2010). pp. 39--46. {ACM} (2010),
  \url{https://doi.org/10.1145/1864708.1864721}

\bibitem{Dietterich1998}
Dietterich, T.G.: {Approximate Statistical Tests for Comparing Supervised
  Classification Learning Algorithms}. Neural Comput.  \textbf{10}(7),
  1895--1923 (1998), \url{https://doi.org/10.1162/089976698300017197}

\bibitem{Dorigo2006}
Dorigo, M., Birattari, M., Stutzle, T.: Ant colony optimization. IEEE
  Computational Intelligence Magazine  \textbf{1}(4),  28--39 (2006).
  \doi{10.1109/MCI.2006.329691}

\bibitem{WorldBankSME}
Group, T.W.B.: {Small and Medium Enterprises (SMEs) Finance} (2022),
  \url{https://www.worldbank.org/en/topic/smefinance}, last accessed:
  2022-10-04

\bibitem{10.1145/3397271.3401060}
Hansen, C., Hansen, C., Simonsen, J.G., Alstrup, S., Lioma, C.: Content-aware
  neural hashing for cold-start recommendation. In: Proceedings of the 43rd
  International ACM SIGIR Conference on Research and Development in Information
  Retrieval. p. 971–980. SIGIR '20, Association for Computing Machinery, New
  York, NY, USA (2020). \doi{10.1145/3397271.3401060},
  \url{https://doi.org/10.1145/3397271.3401060}

\bibitem{conf/edm/HansenHHAL17}
Hansen, C., Hansen, C., Hjuler, N., Alstrup, S., Lioma, C.: Sequence modelling
  for analysing student interaction with educational systems. In: Hu, X.,
  Barnes, T., Hershkovitz, A., Paquette, L. (eds.) EDM. International
  Educational Data Mining Society (IEDMS) (2017),
  \url{http://dblp.uni-trier.de/db/conf/edm/edm2017.html#HansenHHAL17}

\bibitem{10.1145/3442381.3450011}
Hansen, C., Hansen, C., Simonsen, J.G., Lioma, C.: Projected hamming
  dissimilarity for bit-level importance coding in collaborative filtering. In:
  Proceedings of the Web Conference 2021. p. 261–269. WWW '21, Association
  for Computing Machinery, New York, NY, USA (2021).
  \doi{10.1145/3442381.3450011}, \url{https://doi.org/10.1145/3442381.3450011}

\bibitem{HarperEtAl2016}
Harper, F.M., Konstan, J.A.: {The MovieLens Datasets: History and Context}.
  {ACM} Trans. Interact. Intell. Syst.  \textbf{5}(4),  19:1--19:19 (2016).
  \doi{10.1145/2827872}, \url{https://doi.org/10.1145/2827872}

\bibitem{He2017}
He, X., Liao, L., Zhang, H., Nie, L., Hu, X., Chua, T.: Neural collaborative
  filtering. In: Barrett, R., Cummings, R., Agichtein, E., Gabrilovich, E.
  (eds.) Proceedings of the 26th International Conference on World Wide Web,
  ({WWW} 2017). pp. 173--182. {ACM} (2017),
  \url{https://doi.org/10.1145/3038912.3052569}

\bibitem{genetic_algorithm}
Holland, J.H.: Genetic algorithms. Scientific American  \textbf{267}(1),
  66--73 (1992), \url{http://www.jstor.org/stable/24939139}

\bibitem{InozemtsevaEtAl2014}
Inozemtseva, L., Holmes, R., Walker, R.J.: Recommendation systems in-the-small.
  In: Robillard, M.P., Maalej, W., Walker, R.J., Zimmermann, T. (eds.)
  Recommendation Systems in Software Engineering, pp. 77--92. Springer (2014),
  \url{https://doi.org/10.1007/978-3-642-45135-5\_4}

\bibitem{KaminskasEtAl2015}
Kaminskas, M., Bridge, D., Foping, F.S., Roche, D.: Product recommendation for
  small-scale retailers. In: Stuckenschmidt, H., Jannach, D. (eds.) Proceedings
  of the 16th International Conference on Electronic Commerce and Web
  Technologies, (EC-Web 2015). Lecture Notes in Business Information
  Processing, vol.~239, pp. 17--29. Springer (2015),
  \url{https://doi.org/10.1007/978-3-319-27729-5\_2}

\bibitem{KaminskasEtAl2017}
Kaminskas, M., Bridge, D., Foping, F.S., Roche, D.: Product-seeded and
  basket-seeded recommendations for small-scale retailers. J. Data Semant.
  \textbf{6}(1),  3--14 (2017), \url{https://doi.org/10.1007/s13740-016-0058-3}

\bibitem{Kennedy1995}
Kennedy, J., Eberhart, R.: Particle swarm optimization. In: Proceedings of
  ICNN'95 - International Conference on Neural Networks. vol.~4, pp. 1942--1948
  vol.4 (1995). \doi{10.1109/ICNN.1995.488968}

\bibitem{kutner2005}
Kutner, M., , Nachtsheim, C.J., Neter, J., Li, W., et~al.: {Applied Linear
  Statistical Models}. McGraw-Hill, Irwin (2005)

\bibitem{Kuzelewska2020}
Kuzelewska, U.: {Effect of Dataset Size on Efficiency of Collaborative
  Filtering Recommender Systems with Multi-clustering as a Neighbourhood
  Identification Strategy}. In: Krzhizhanovskaya, V.V., Z{\'{a}}vodszky, G.,
  Lees, M.H., Dongarra, J.J., Sloot, P.M.A., Brissos, S., Teixeira, J. (eds.)
  Proceedings of the 20th International Conference on Computational Science
  ({ICCS} 2020), Part {III}. Lecture Notes in Computer Science, vol. 12139, pp.
  342--354. Springer (2020),
  \url{https://doi.org/10.1007/978-3-030-50420-5\_25}

\bibitem{LatifiEtAl2021}
Latifi, S., Mauro, N., Jannach, D.: {Session-aware Recommendation: A Surprising
  Quest for the State-of-the-art}. Inf. Sci.  \textbf{573},  291--315 (2021),
  \url{https://doi.org/10.1016/j.ins.2021.05.048}

\bibitem{LeeEtAl2021}
Lee, D., Kang, S., Ju, H., Park, C., Yu, H.: {Bootstrapping User and Item
  Representations for One-Class Collaborative Filtering}. In: Diaz, F., Shah,
  C., Suel, T., Castells, P., Jones, R., Sakai, T. (eds.) Proceedings of the
  44th International {ACM} Conference on Research and Development in
  Information Retrieval, (SIGIR 2021). pp. 1513--1522. {ACM} (2021),
  \url{https://doi.org/10.1145/3404835.3462935}

\bibitem{Lee2013}
Lee, S., Park, S., Kahng, M., Lee, S.: {PathRank: Ranking Nodes on a
  Heterogeneous Graph for Flexible Hybrid Recommender Systems}. Expert Systems
  with Applications  \textbf{40}(2),  684--697 (2013),
  \url{https://doi.org/10.1016/j.eswa.2012.08.004}

\bibitem{LeeEtAl2009}
Lee, Y., Cheng, T., Lan, C., Wei, C., Hu, P.J.: Overcoming small-size training
  set problem in content-based recommendation: a collaboration-based training
  set expansion approach. In: Chau, P.Y.K., Lyytinen, K., Wei, C., Yang, C.C.,
  Lin, F. (eds.) Proceedings of the 11th International Conference on Electronic
  Commerce, (ICEC 2009). pp. 99--106. {ACM} (2009),
  \url{https://doi.org/10.1145/1593254.1593268}

\bibitem{LudewigEtAl2021}
Ludewig, M., Mauro, N., Latifi, S., Jannach, D.: {Empirical Analysis of
  Session-based Recommendation Algorithms}. User Model. User Adapt. Interact.
  \textbf{31}(1),  149--181 (2021),
  \url{https://doi.org/10.1007/s11257-020-09277-1}

\bibitem{SmallDataProblems}
Ng, A.Y.T.: {Why AI Projects Fail, Part 4: Small Data} (2019),
  \url{https://www.deeplearning.ai/the-batch/why-ai-projects-fail-part-4-small-data/},
  last accessed: 2022-10-04

\bibitem{Odili2018}
Odili, J.: The dawn of metaheuristic algorithms. International Journal of
  Software Engineering and Computer Systems  \textbf{4},  49--61 (08 2018).
  \doi{10.15282/ijsecs.4.2.2018.4.0048}

\bibitem{Page1999}
Page, L., Brin, S., Motwani, R., Winograd, T.: The pagerank citation ranking :
  Bringing order to the web. In: WWW 1999 (1999)

\bibitem{PanEtAl2022}
Pan, X., Li, M., Zhang, J., Yu, K., Wen, H., Wang, L., Mao, C., Cao, B.:
  Metacvr: Conversion rate prediction via meta learning in small-scale
  recommendation scenarios. In: Amig{\'{o}}, E., Castells, P., Gonzalo, J.,
  Carterette, B., Culpepper, J.S., Kazai, G. (eds.) Proceedings of the 45th
  International {ACM} {SIGIR} Conference on Research and Development in
  Information Retrieval, (SIGIR 2022). pp. 2110--2114. {ACM} (2022),
  \url{https://doi.org/10.1145/3477495.3531733}

\bibitem{RaziperchikolaeiEtAl2021}
Raziperchikolaei, R., Liang, G., Chung, Y.: {Shared Neural Item Representations
  for Completely Cold Start Problem}. In: Pamp{\'{\i}}n, H.J.C., Larson, M.A.,
  Willemsen, M.C., Konstan, J.A., McAuley, J.J., Garcia{-}Gathright, J.,
  Huurnink, B., Oldridge, E. (eds.) Proceedings of the 15th {ACM} Conference on
  Recommender Systems, (RecSys 2021). pp. 422--431. {ACM} (2021),
  \url{https://doi.org/10.1145/3460231.3474228}

\bibitem{Salha-GalvanEtAl2021}
Salha{-}Galvan, G., Hennequin, R., Chapus, B., Tran, V., Vazirgiannis, M.:
  {Cold Start Similar Artists Ranking with Gravity-Inspired Graph
  Autoencoders}. In: Pamp{\'{\i}}n, H.J.C., Larson, M.A., Willemsen, M.C.,
  Konstan, J.A., McAuley, J.J., Garcia{-}Gathright, J., Huurnink, B., Oldridge,
  E. (eds.) Proceedings of the 15th {ACM} Conference on Recommender Systems,
  (RecSys 2021). pp. 443--452. {ACM} (2021),
  \url{https://doi.org/10.1145/3460231.3474252}

\bibitem{SankarEtAl2021}
Sankar, A., Wang, J., Krishnan, A., Sundaram, H.: {ProtoCF: Prototypical
  Collaborative Filtering for Few-shot Recommendation}. In: Pamp{\'{\i}}n,
  H.J.C., Larson, M.A., Willemsen, M.C., Konstan, J.A., McAuley, J.J.,
  Garcia{-}Gathright, J., Huurnink, B., Oldridge, E. (eds.) Proceedings of the
  15th {ACM} Conference on Recommender Systems, (RecSys 2021). pp. 166--175.
  {ACM} (2021), \url{https://doi.org/10.1145/3460231.3474268}

\bibitem{SchnabelAndBennett2020}
Schnabel, T., Bennett, P.N.: Debiasing item-to-item recommendations with small
  annotated datasets. In: Santos, R.L.T., Marinho, L.B., Daly, E.M., Chen, L.,
  Falk, K., Koenigstein, N., de~Moura, E.S. (eds.) Proceedings of the 14th
  {ACM} Conference on Recommender Systems, (RecSys 2020). pp. 73--81. {ACM}
  (2020), \url{https://doi.org/10.1145/3383313.3412265}

\bibitem{ShuaiEtAl2022}
Shuai, J., Zhang, K., Wu, L., Sun, P., Hong, R., Wang, M., Li, Y.: Proceedings
  of the 45th international {ACM} {SIGIR} conference on research and
  development in information retrieval, (sigir 2022). pp. 1283--1293. {ACM}
  (2022), \url{https://doi.org/10.1145/3477495.3531927}

\bibitem{StrickrothAndPinkwart2012}
Strickroth, S., Pinkwart, N.: High quality recommendations for small
  communities: the case of a regional parent network. In: Cunningham, P.,
  Hurley, N.J., Guy, I., Anand, S.S. (eds.) Proceedings of the 6th {ACM}
  Conference on Recommender Systems, (RecSys 2012). pp. 107--114. {ACM} (2012),
  \url{https://doi.org/10.1145/2365952.2365976}

\bibitem{SunXZEtAl2021}
Sun, H., Xu, J., Zheng, K., Zhao, P., Chao, P., Zhou, X.: {MFNP:} {A}
  meta-optimized model for few-shot next {POI} recommendation. In: Zhou, Z.
  (ed.) Proceedings of the 30th International Joint Conference on Artificial
  Intelligence, ({IJCAI} 2021). pp. 3017--3023. ijcai.org (2021),
  \url{https://doi.org/10.24963/ijcai.2021/415}

\bibitem{SunEtAl2021}
Sun, X., Shi, T., Gao, X., Kang, Y., Chen, G.: {FORM: Follow the Online
  Regularized Meta-Leader for Cold-Start Recommendation}. In: Diaz, F., Shah,
  C., Suel, T., Castells, P., Jones, R., Sakai, T. (eds.) Proceedings of the
  44th International {ACM} Conference on Research and Development in
  Information Retrieval, (SIGIR 2021). pp. 1177--1186. {ACM} (2021),
  \url{https://doi.org/10.1145/3404835.3462831}

\bibitem{SunAndHan2012}
Sun, Y., Han, J.: {Mining Heterogeneous Information Networks: Principles and
  Methodologies}. Synthesis Lectures on Data Mining and Knowledge Discovery,
  Morgan {\&} Claypool Publishers (2012),
  \url{https://doi.org/10.2200/S00433ED1V01Y201207DMK005}

\bibitem{SunEtAl2020}
Sun, Z., Yu, D., Fang, H., Yang, J., Qu, X., Zhang, J., Geng, C.: {Are We
  Evaluating Rigorously? Benchmarking Recommendation for Reproducible
  Evaluation and Fair Comparison}. In: Santos, R.L.T., Marinho, L.B., Daly,
  E.M., Chen, L., Falk, K., Koenigstein, N., de~Moura, E.S. (eds.) Proceedings
  of the 14th {ACM} Conference on Recommender Systems, (RecSys 2020). pp.
  23--32. {ACM} (2020), \url{https://doi.org/10.1145/3383313.3412489}

\bibitem{VolkovsEtAl2017}
Volkovs, M., Yu, G.W., Poutanen, T.: {DropoutNet: Addressing Cold Start in
  Recommender Systems}. In: Guyon, I., von Luxburg, U., Bengio, S., Wallach,
  H.M., Fergus, R., Vishwanathan, S.V.N., Garnett, R. (eds.) Proceedings of the
  30th Annual Conference on Neural Information Processing Systems (NeurIPS
  2017). pp. 4957--4966 (2017),
  \url{https://proceedings.neurips.cc/paper/2017/hash/dbd22ba3bd0df8f385bdac3e9f8be207-Abstract.html}

\bibitem{WangQinyongEtAl2019}
Wang, Q., Yin, H., Wang, H., Nguyen, Q.V.H., Huang, Z., Cui, L.: {Enhancing
  Collaborative Filtering with Generative Augmentation}. In: Teredesai, A.,
  Kumar, V., Li, Y., Rosales, R., Terzi, E., Karypis, G. (eds.) Proceedings of
  the 25th {ACM} International Conference on Knowledge Discovery and Data
  Mining, (SIGKDD 2019). pp. 548--556. {ACM} (2019),
  \url{https://doi.org/10.1145/3292500.3330873}

\bibitem{ShoujinWangEtAl2021}
Wang, S., Hu, L., Wang, Y., He, X., Sheng, Q.Z., Orgun, M.A., Cao, L., Ricci,
  F., Yu, P.S.: {Graph Learning based Recommender Systems: A Review}. In: Zhou,
  Z. (ed.) Proceedings of the 30th International Joint Conference on Artificial
  Intelligence, ({IJCAI} 2021). pp. 4644--4652. ijcai.org (2021),
  \url{https://doi.org/10.24963/ijcai.2021/630}

\bibitem{WangEtAl2021}
Wang, S., Zhang, K., Wu, L., Ma, H., Hong, R., Wang, M.: {Privileged Graph
  Distillation for Cold Start Recommendation}. In: Diaz, F., Shah, C., Suel,
  T., Castells, P., Jones, R., Sakai, T. (eds.) Proceedings of the 44th
  International {ACM} Conference on Research and Development in Information
  Retrieval, (SIGIR 2021). pp. 1187--1196. {ACM} (2021),
  \url{https://doi.org/10.1145/3404835.3462929}

\bibitem{WangEtAl2019}
Wang, X., He, X., Cao, Y., Liu, M., Chua, T.: {KGAT: Knowledge Graph Attention
  Network for Recommendation}. In: Teredesai, A., Kumar, V., Li, Y., Rosales,
  R., Terzi, E., Karypis, G. (eds.) Proceedings of the 25th {ACM} International
  Conference on Knowledge Discovery {\&} Data Mining, (SIGKDD 2017). pp.
  950--958. {ACM} (2019), \url{https://doi.org/10.1145/3292500.3330989}

\bibitem{Wang2019}
Wang, X., He, X., Wang, M., Feng, F., Chua, T.S.: Neural graph collaborative
  filtering. In: Proceedings of the 42nd International ACM SIGIR Conference on
  Research and Development in Information Retrieval. p. 165–174. SIGIR'19,
  Association for Computing Machinery, New York, NY, USA (2019).
  \doi{10.1145/3331184.3331267}, \url{https://doi.org/10.1145/3331184.3331267}

\bibitem{WuEtAl2021}
Wu, J., Wang, X., Feng, F., He, X., Chen, L., Lian, J., Xie, X.:
  {Self-supervised Graph Learning for Recommendation}. In: Diaz, F., Shah, C.,
  Suel, T., Castells, P., Jones, R., Sakai, T. (eds.) Proceedings of the 44th
  International {ACM} Conference on Research and Development in Information
  Retrieval, (SIGIR 2021). pp. 726--735. {ACM} (2021),
  \url{https://doi.org/10.1145/3404835.3462862}

\bibitem{WuEtAl2022}
Wu, J., Xie, Z., Yu, T., Zhao, H., Zhang, R., Li, S.: Dynamics-aware adaptation
  for reinforcement learning based cross-domain interactive recommendation. In:
  Amig{\'{o}}, E., Castells, P., Gonzalo, J., Carterette, B., Culpepper, J.S.,
  Kazai, G. (eds.) Proceedings of the 45th International {ACM} {SIGIR}
  Conference on Research and Development in Information Retrieval, (SIGIR
  2022). pp. 290--300. {ACM} (2022),
  \url{https://doi.org/10.1145/3477495.3531969}

\bibitem{XiaEtAl2022}
Xia, L., Huang, C., Xu, Y., Zhao, J., Yin, D., Huang, J.X.: Hypergraph
  contrastive collaborative filtering. In: Amig{\'{o}}, E., Castells, P.,
  Gonzalo, J., Carterette, B., Culpepper, J.S., Kazai, G. (eds.) Proceedings of
  the 45th International {ACM} {SIGIR} Conference on Research and Development
  in Information Retrieval, (SIGIR 2022). pp. 70--79. {ACM} (2022),
  \url{https://doi.org/10.1145/3477495.3532058}

\bibitem{Xiang2010}
Xiang, L., Yuan, Q., Zhao, S., Chen, L., Zhang, X., Yang, Q., Sun, J.:
  {Temporal Recommendation on Graphs via Long- and Short-term Preference
  Fusion}. In: Rao, B., Krishnapuram, B., Tomkins, A., Yang, Q. (eds.)
  Proceedings of the 16th {ACM} International Conference on Knowledge Discovery
  {\&} Data Mining, (SIGKDD 2010). pp. 723--732. {ACM} (2010),
  \url{https://doi.org/10.1145/1835804.1835896}

\bibitem{Yao2013}
Yao, W., He, J., Huang, G., Cao, J., Zhang, Y.: {Personalized Recommendation on
  Multi-Layer Context Graph}. In: Lin, X., Manolopoulos, Y., Srivastava, D.,
  Huang, G. (eds.) {Proceedings of the 14th International Conference on Web
  Information Systems Engineering (WISE 2013), Part I}. Lecture Notes in
  Computer Science, vol.~8180, pp. 135--148. Springer (2013),
  \url{https://doi.org/10.1007/978-3-642-41230-1\_12}

\bibitem{Yu2014}
Yu, X., Ren, X., Sun, Y., Gu, Q., Sturt, B., Khandelwal, U., Norick, B., Han,
  J.: {Personalized Entity Recommendation: a Heterogeneous Information Network
  Approach}. In: Carterette, B., Diaz, F., Castillo, C., Metzler, D. (eds.)
  Proceedings of the 7th {ACM} International Conference on Web Search and Data
  Mining, (WSDM 2014). pp. 283--292. {ACM} (2014),
  \url{https://doi.org/10.1145/2556195.2556259}

\end{thebibliography}
%




\end{document}